\documentclass[onecolumn,letterpaper,amsmath,amssymb,floatfix,aps,superscriptaddress]{revtex4}
\usepackage{subfigure}
\usepackage{color}

\usepackage{graphicx}
\usepackage{bm}  
\usepackage{epsfig}
\usepackage{mathtools}

\begin{document}

\title{Tensorial conservation law for nematic polymers}

\date{\today}

\author{Daniel Sven\v sek}
\affiliation{Department of Physics, Faculty of Mathematics and Physics, University of Ljubljana, Jadranska 19, SI-1111 Ljubljana, Slovenia}

\author{Gregory M. Grason}
\affiliation{Department of Polymer Science and Engineering, University of Massachusetts, Amherst, MA 01003, USA}

\author{Rudolf Podgornik}
\affiliation{Department of Physics, Faculty of Mathematics and Physics, University of Ljubljana, Jadranska 19, SI-1111 Ljubljana, Slovenia}
\affiliation{Department of Theoretical Physics, J. Stefan Institute, Jamova 39,  SI-1111 Ljubljana, Slovenia}
\affiliation{Department of Physics,  University of Massachusetts, Amherst MA 01003, USA}

\begin{abstract}
We derive the ``conservation law" for nematic polymers in tensorial form valid for quadrupolar orientational order in contradistinction to the conservation law
in the case of polar orientational order. Due to microscopic differences in the coupling between the orientational field deformations and the density variations for polar and quadrupolar order,  we find that respective order parameters satisfy fundamentally distinct constraints.  Being necessarily scalar in its form, the tensorial conservation law is obtained straightforwardly from the gradients of the polymer nematic tensor field and connects the spatial variation of this tensor field with density variations. We analyze the differences between the polar and the tensorial forms of the conservation law, present some explicit orientational fields that satisfy this new constraint and discuss the role of singular ``hairpins'', which do not affect local quadrupolar order of polymer nematics, but nevertheless influence its gradients.
\end{abstract}


\maketitle


\section{Introduction and background}
\label{background}

\noindent
In polymer nematics, deformations of orientational order are inextricably linked to density.  Splay deformations necessarily introduce local changes in polymer density, which become progressively more expensive as chain length grows \cite{conteq, meyer1,meyer,kamien,Kamien2,selinger_bruinsma_pra,kamien_toner}. In the continuous limit of long chains this coupling between the polymer density and orientational fields is described by an analogue of the continuity equation for the nematic director field $\bf n$ \cite{svensek-podgornik,svensek-podgornik_chiral}, 
\begin{equation}
	\nabla\cdot(\rho_s{\bf n}) = 0,
	\label{continuity_basic}
\end{equation}
where $\rho_s$ is the areal number density of chains crossing the plane perpendicular to the director field \cite{meyer}. 
The constraint imposed by the polar nematic order can be interpreted as the continuity condition for a ``polymer current density''  ${\bf j} = \rho_s {\bf n}$ and the only difference w.r.t. the usual continuity equation is that in this case $\bf j$ does not describe a rate (there is no time derivative involved in its definition), i.e., we are observing the number of chains perforating the perpendicular plane rather than the number of particles crossing it per unit time.  The analogy comes fully into life if one relaxes the condition $|{\bf n}|=1$ and takes into account the variable degree of nematic order. Then one can define the complete {\em nematic polymer current density} \cite{svensek-podgornik,svensek-podgornik_chiral} ${\bf j}({\bf x}) = \rho({\bf x})\, \ell_0\, {\bf a}({\bf x})$
that obeys the conservation law
\begin{equation}
	\nabla\cdot{\bf j} = \rho^+ -\rho^-,
	\label{current_continuity}
\end{equation}
with $\rho({\bf x})$ the volume number density of monomers (or any arbitrarily defined chain segments), $\ell_0$ the corresponding monomer length, and $\rho^\pm({\bf x})$ the volume number density of the beginnings ($\rho^+$) and the ends ($\rho^-$) of the chains. The non-unit nematic vector order parameter $\bf a$ is defined as the average of the monomer unit vectors ${\bf d}^\alpha$, $	{\bf a} = \langle {\bf d}^\alpha\rangle = \langle{\bf d}^\alpha\cdot {\bf n}\rangle\, {\bf n} \label{a_brief} $, where $\bf n$ is a unit polar nematic director and $\langle{\bf d}^\alpha\cdot {\bf n}\rangle$ is the degree of polar order. 
The areal number density of perforating chains $\rho_s$ in Eq.~(\ref{continuity_basic}) is thus $\rho_s = \rho \ell_0 \vert{\bf a}\vert$.
The continuity equation (\ref{current_continuity}) for the polymer current presents a coupling of the director splay deformation and the variation of the density, which in the infinite chain limit $\rho^\pm=0$ strictly follows the constraint
\begin{equation}
	\rho\, \nabla\cdot{\bf a} + {\bf a}\cdot\nabla\rho = 0.
	\label{current_continuity_nosources}	
\end{equation}
Note that this constraint involves also the variations of the degree of order $\vert{\bf a}\vert$, besides the splay deformation of the director field.

\begin{figure}[htb]
\begin{center}
	\mbox{
		\subfigure[\,+1/2]{\includegraphics[width=30mm]{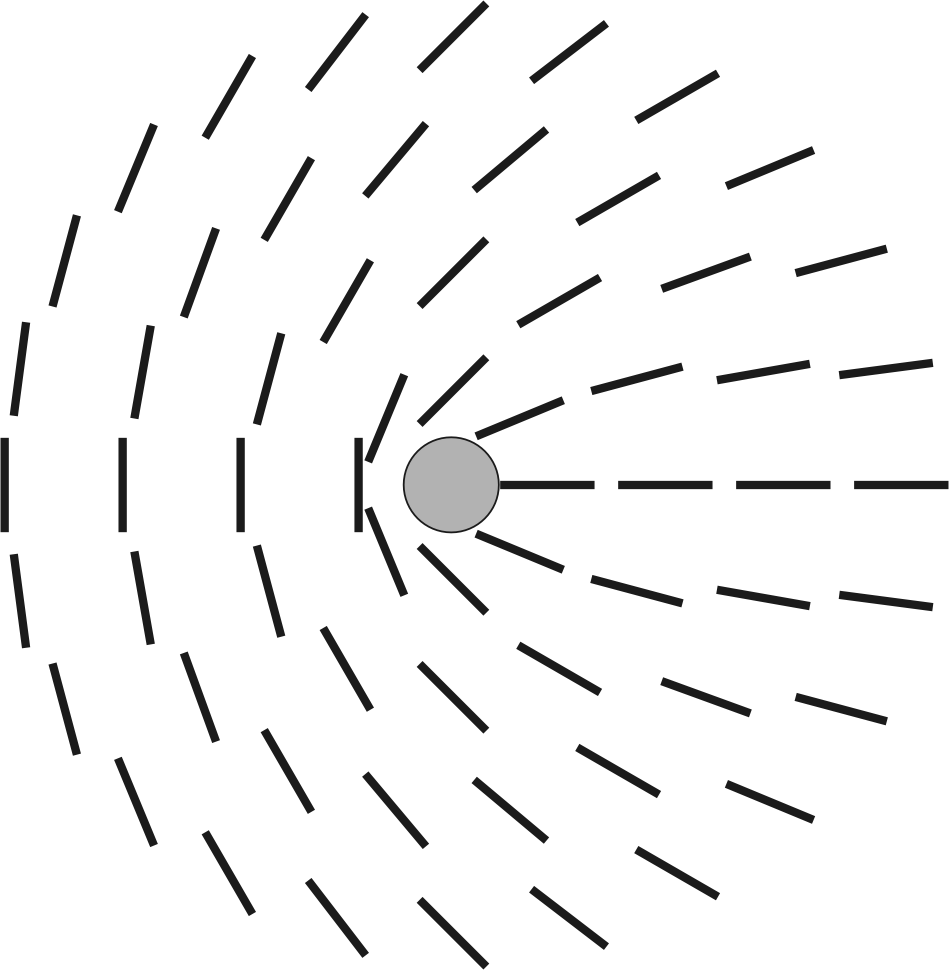}}\hspace{5mm}  
		\subfigure[\,+1]{\includegraphics[width=30mm]{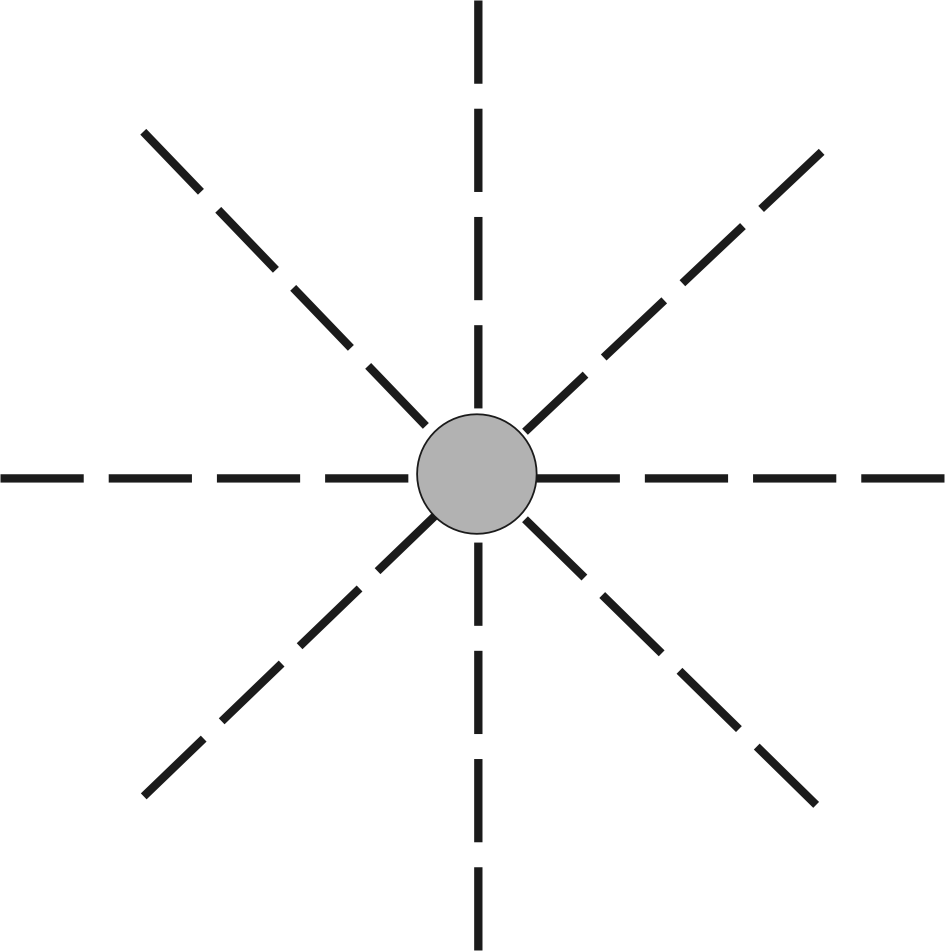}} 
	}
	\caption{Director field (schematic) of the +1/2 and +1 disclinations. A cross section perpendicular to the disclination line of the wedge type is shown; (a) is possible only in the quadrupolar phase.
	\label{defects}
	}
\end{center}
\end{figure}

\begin{figure}
\begin{center}
	\mbox{
		\subfigure[\, no hairpins]{\includegraphics[height=20mm]{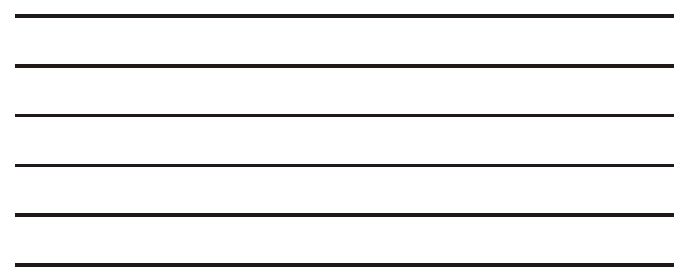}}\hspace{2mm}  
		\subfigure[\, hairpins]{\includegraphics[height=20mm]{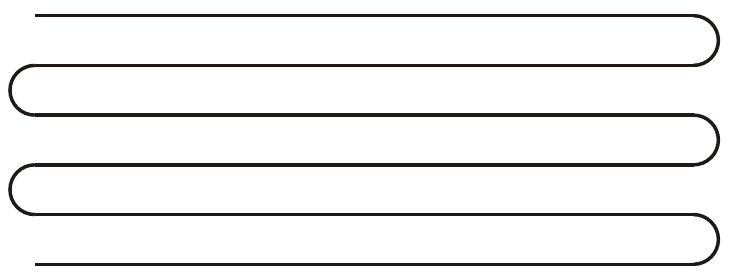}} 
	}
	\caption{A schematic representation of perfectly ordered chains with and without hairpin turns: neglecting the U-turn regions in (b), the orientational order is the same in both cases, if the chain possesses inversion symmetry, while the configurations are physically different with respect to the connectivity.
	\label{connectivity_polarity}
	}
\end{center}
\end{figure}

While previous theoretical treatments have focused on the structure of polar chain order and the consequent constraints for the vector order parameter, in this paper we consider the consequences of microscopic chain constructure for the quadrupolar order of a true polymer nematic, which respects the symmetry of local reversal of chain orientation (${\bf d}^\alpha \to - {\bf d}^\alpha$).
Unlike the polar state,  nematic is described by a tensorial order parameter $\sf Q$~\cite{degennes_prost}
\begin{equation}
	Q_{ij} = {3\over 2}\left(\langle d^\alpha_i d^\alpha_j\rangle - {{1\over 3}}\delta_{ij}\right).
	\label{Q_brief}
\end{equation}
The most obvious difference between the polar and the quadrupolar phases is the existence of $\pm 1/2$ disclinations in the latter, Fig.~\ref{defects}.   
Even in the global or local absence of topological defects where the $\sf Q$-tensor is uniaxial and expressed by the polar nematic director $\bf n$,
\begin{equation}
	Q_{ij}={3S\over 2}\left(n_i n_j - {1\over 3}\delta_{ij}\right),\quad  
	S = {3\over 2}\left(\langle ({\bf d}^\alpha\cdot{\bf n})^2\rangle - {1\over 3}\right),
	\label{Q_uniaxial}
\end{equation}
there exist microscopic differences between the polar and a quadrupolar phases beyond the ${\bf n}\leftrightarrow -{\bf n}$ symmetry of the latter.  Namely, for a polymer the notion of polarity is slightly more delicate, as only the orientational part of the ordering is captured by the nematic order parameter, but not the chain connectivity.  When the underlying building element itself (the monomer) is nonpolar and thus the phase cannot exhibit any polar orientational order by construction, the underlying chain connectivity may exhibit a rich spectrum of states distinguished in terms of microscopic configurations of  ``hairpins'', Fig.~\ref{connectivity_polarity}.
In the quadrupolar phase the hairpins can be present without destroying or altering the orientational order, but certainly not in the polar phase~\cite{kamien}.  Hairpins are not defects of the average mesoscopic orientational field, but rather microscopic wiggles not captured by the mesoscopic order parameter. 

Hence, even in the simple uniaxial case in which the quadrupolar and polar phases look macroscopically alike, one anticipates that the coupling of the orientational field deformations and the density variations will be different.  As the nematic polymer current density and the continuity equation it satisfies (\ref{current_continuity}) are vectorial in nature, they cannot be applied directly to the quadrupolar case, where the vector $\bf a$ cannot be defined. The question thus arises what is the correct form of the ordering constraint valid for polymer chains with inversion symmetry and the flexibility to make hairpin turns. Moreover, we are led to ask what operators, if not the nematic order parameter tensor itself, carry information about the distribution of microscopic hairpins, whose energy cost is clearly not encoded in smooth distortions of the mean orientation ${\bf n}$.  In this paper, we derive the constraint that must be satisfied by the nematic order parameter of chain systems exhibiting states of quadrupolar orientational order.  

The remainder of the paper is organized as follows.  In Sec.~\ref{derivation} we present the conservation law obeyed by the nematic order parameter for polymers, and in Sec.~\ref{hairpins}, we analyze how hairpins that do not contribute to the local value of the quadrupolar order parameter nevertheless influence its gradients.  In Sec.~\ref{vecvstens} we analyze example textures to highlight differences in the mesoscopic textures exhibited by conserved polar and quadrupolar polymer phases and conclude with a brief discussion of the consequences of the nematic conservation law.

\section{Derivation of a tensorial conservation law}
\label{derivation}

\noindent
For the sake of comparison we begin with a derivation of the vectorial conservation law for polar polymer order.
First we define a collection of microscopic fields, where the integral is performed over the polymer contour ${\bf x}(s)$:
the microscopic polymer areal density field 
\begin{equation}
	\rho^{mic}({\bf x}) =  \int_{{\bf x}(s)} ds ~\delta({\bf x} - {\bf x}(s)), 
	\label{rho_mic}
\end{equation}
the microscopic polymer nematic vector field
\begin{equation}
	j_i^{mic}({\bf x}) =  \int_{{\bf x}(s)} ds~\delta({\bf x} - {\bf x}(s)) ~t_i({\bf x}(s)) ,  \qquad {\rm with} \quad t_i({\bf x}(s)) = \frac{d x_i(s)}{ds},
	\label{j_mic}
\end{equation}
the microscopic polymer nematic tensor field 
\begin{equation}
	\tilde{J}_{ij}^{mic}({\bf x}) = \int_{{\bf x}(s)} ds ~\delta({\bf x} - {\bf x}(s))\, {\textstyle{3\over 2}}t_i({\bf x}(s)) t_j({\bf x}(s)), 
	\label{J_mic}
\end{equation}
and its traceless version
\begin{equation}
	J_{ij}^{mic}({\bf x}) =  \int_{{\bf x}(s)} ds ~\delta({\bf x} - {\bf x}(s))\, 
				{\textstyle{3\over 2}}\left[t_i({\bf x}(s)) t_j({\bf x}(s))-{\textstyle{1\over 3}}\delta_{ij}\right] =
				\tilde{J}^{mic}_{ij}({\bf x}) - \textstyle{1\over 2}\delta_{ij}\rho^{mic}({\bf x}).
	\label{J_traceless_mic}
\end{equation}

Let ${\bf f}({\bf x}(s))$ be any (scalar, vector, or tensor) function of the position on the polymer ${\bf x}(s)$, and ${\bf F}^{mic}$ the associated microscopic field,
\begin{equation}
	{\bf F}^{mic}({\bf x}) = \int_{{\bf x}(s)} ds ~\delta({\bf x} - {\bf x}(s)) ~{\bf f}({\bf x}(s)).
\end{equation}
Coarse graining ${\bf F}^{mic}({\bf x})$ to the mesoscopic volume $V$ centered at $\bf x$ (denoted by $\overbracket{\phantom{F^{mic}}}$~) gives the corresponding mesoscopic field
\begin{equation}
	{\bf F}({\bf x}) = \overbracket{{\bf F}^{mic}}({\bf x})=
			   {1\over V}\int_{V({\bf x})} d^3 x' ~{\bf F}^{mic}({\bf x}') =
			   {1\over V} \int_{{\bf x}(s)\in V({\bf x})} ds ~{\bf f}({\bf x}(s)) =
			   {L({\bf x})\over V} {1\over L({\bf x})}\int_{{\bf x}(s)\in V({\bf x})} ds ~{\bf f}({\bf x}(s)),
\end{equation}
where $L({\bf x})=\int_{{\bf x}(s)\in V({\bf x})}~ds \equiv N({\bf x})\ell_0$ is the total length of the polymer within the volume $V$, which can be expressed in terms of the segment length $\ell_0$ and the number of segments $N$ within the volume. 
Hence, in the continuum spirit the macroscopic field can be written as
\begin{equation}
	{\bf F}({\bf x}) = \rho({\bf x}) \ell_0~ \bar{{\bf f}}({\bf x}), \qquad {\rm with} \qquad \rho({\bf x}) = {N({\bf x})\over V},
	\label{F_meso}
\end{equation}
where $ \rho({\bf x})$ is the (mesoscopic) number density of segments and
\begin{equation}
	\bar{{\bf f}}({\bf x}) = {1\over L({\bf x})}\int_{{\bf x}(s)\in V({\bf x})} ds ~{\bf f}({\bf x}(s))
\end{equation}
is the (mesoscopic) average of ${\bf f}({\bf x}(s))$ (the nematic vector $\bf a$ or nematic tensor ${\sf Q}$ order parameters in what follows). For ${\bf f}({\bf x}(s))=1$ one obtains from Eq.~(\ref{F_meso}) correctly that the mesoscopic version of the microscopic density as defined in Eq.~(\ref{rho_mic}) is $\rho\ell_0$. In the vector case, Eq.~(\ref{j_mic}), one gets the mesoscopic polymer current density
\begin{equation}
	{\bf j} = \overbracket{{\bf j}^{mic}}=\rho \ell_0 {\bf a},\quad
			{\bf a} = {1\over L({\bf x})}\int_{{\bf x}(s)\in V({\bf x})} ds ~{\bf t}({\bf x}(s))=
					\langle {\bf d}^\alpha\rangle,
	\label{j_meso}
\end{equation}
which is in accord with the definition of the nematic polymer current density in Sec.~\ref{background}.  In the traceless tensor case, Eq.~(\ref{J_traceless_mic}), the mesoscopic field is the following:
\begin{equation}
	{\sf J} = \overbracket{{\sf J}^{mic}}=\rho \ell_0 {\sf Q},\quad
	Q_{ij} = {1\over L({\bf x})}\int_{{\bf x}(s)\in V({\bf x})} ds ~{\textstyle{3\over 2}}\left[t_i({\bf x}(s)) t_j({\bf x}(s))-{\textstyle{1\over 3}}\delta_{ij}\right]={\textstyle{3\over 2}}\left(\langle d^\alpha_i d^\alpha_j\rangle - {\textstyle{1\over 3}}\delta_{ij}\right),
	\label{J_meso}
\end{equation}
where $Q_{ij}$ agrees with the definition of the nematic order tensor in Eq.~(\ref{Q_brief}).

As promised, let us first briefly recapitulate the derivation of the vectorial conservation law. The divergence of the microscopic nematic vector field, Eq.~(\ref{j_mic}), is given by
\begin{eqnarray}
	\partial_i j^{mic}_{i}({\bf x}) &=&  \int_{{\bf x}(s)} ds ~\partial_j \delta({\bf x} - {\bf x}(s)) ~ t_j({\bf x}(s)) = \int_{{\bf x}(s)} ds			~\frac{\partial}{\partial x_j} \delta({\bf x} - {\bf x}(s)) \frac{d x_j(s)}{ds} = - \int_{{\bf x}(s)} ds~\frac{\partial}{\partial x_j(s)} \delta({\bf x} - {\bf x}(s)) \frac{d x_j(s)}{ds}   = \nonumber\\
	&=& - \int_{{\bf x}(s)} ds ~\frac{d}{ds} \delta({\bf x} - {\bf x}(s)) = \delta({\bf x} - {\bf x}(0)) -  \delta({\bf x} - {\bf x}(L)),
	\label{div_jmic}
\end{eqnarray}
where $L$ is the total length of the chain.
As a side remark one can trivially conclude from this that
\begin{equation}
	\int\! d^3 x\, \partial_i j^{mic}_{i}({\bf x}) =  0,
	\label{nfxgejwq}
\end{equation}
since every polymer chain needs to have a beginning and an end.

The corresponding mesoscopic relation for the divergence of the microscopic nematic vector field is obtained by coarse graining Eq.~(\ref{div_jmic}). As the coarse-graining and $\nabla$ commute, i.e., $\overbracket{\partial_i u} = \partial_i\overbracket{u}$ for any function $u({\bf x})$,
%
%
one gets straightforwardly, $\overbracket{\partial_i j^{mic}_{i}} = \partial_i\overbracket{j^{mic}_{i}}=\partial_i j_i(x)$ 
and thus the coarse-grained version of Eq.~(\ref{div_jmic}) is the polymer vector current conservation law, Eq.~(\ref{current_continuity}).

Let us now turn to the tensorial case. We proceed by first taking the following contraction of the microscopic polymer nematic tensor field (\ref{J_mic}):
\begin{eqnarray}
	\partial_i\partial_j \tilde{J}^{mic}_{ij}({\bf x}) &=&   \int_{{\bf x}(s)} ds~ \frac{d x_i(s)}{ds}  \frac{d x_j(s)}{ds}~\frac{\partial^2}{\partial x_i(s)\partial x_j(s)} \delta({\bf x} - {\bf x}(s)) = \int_{{\bf x}(s)} ds~\frac{d^2}{ds^2} \delta({\bf x} - {\bf x}(s)).
\label{hncfskd}
\end{eqnarray}
As a side remark it follows straightforwardly from here that if integrated over the whole volume
\begin{equation}
	\int\! d^3 x\, \partial_i\partial_j \tilde{J}^{mic}_{ij}({\bf x}) =  0,
\end{equation}
in complete analogy with Eq.~(\ref{nfxgejwq}).
Rewriting Eq.~(\ref{hncfskd}) as
\begin{eqnarray}
	\partial_i\partial_j \tilde{J}^{mic}_{ij}({\bf x}) &=&  
	\frac{d }{ds}  ~\delta({\bf x} - {\bf x}(s))\vert_{s = L} - \frac{d }{ds}  ~\delta({\bf x} - {\bf x}(s))\vert_{s = 0},
\label{hncfskd1}
\end{eqnarray}
where the integral over the length of the chain goes from $0$ to $L$, we furthermore decompose
\begin{equation}
	\frac{d }{ds}  ~\delta({\bf x} - {\bf x}(s)) = -\frac{d x_i(s)}{ds} \frac{\partial}{\partial x_i} \delta({\bf x} - {\bf x}(s)),
\end{equation}
which leads straightforwardly to
\begin{equation}
	\partial_i\partial_j \tilde{J}^{mic}_{ij}({\bf x})  = - \nabla\cdot \left[ \frac{d {\bf x}}{ds}(L)  \delta({\bf x} - {\bf x}(L))\right] + \nabla\cdot \left[ \frac{d {\bf x}}{ds}(0)  \delta({\bf x} - {\bf x}(0))\right].
\end{equation}
By defining the difference of chain orientation at free ends as
\begin{equation}
	\delta {\bf t} ({\bf x} ) \equiv {\bf t}(0)\delta({\bf x} - {\bf x}(0)) - {\bf t}(L)  \delta({\bf x} - {\bf x}(L)) ,
\end{equation}
we obtain
\begin{equation}
	\partial_i\partial_j \tilde{J}^{mic}_{ij}({\bf x})  = \nabla\cdot \delta {\bf t} ({\bf x}).
	\label{J_mic_conservation}
\end{equation}
Upon coarse graining
\begin{eqnarray}
	{1\over V}\int_{V({\bf x})} d^3 x' ~\delta {\bf t} ({\bf x}') &=&	{\bf g}({\bf x}),
\end{eqnarray}
where ${\bf g}$ is the difference of densities of beginnings and ends of chain tangents according to the above convention.  If the system has 
local non-zero polar order of the free end tangents, 
${\bf g}$ presents an additional macroscopic variable in the tensor case, similar to the additional variable in the vector case, $\rho^\pm=\rho^+ - \rho^-$ while for the case of quadrupolar order, ${\bf t} \to - {\bf t}$ symmetry demands that ${\bf g} =0$. 

To make the connection with the nematic tensor order parameter $\sf Q$, Eqs.~(\ref{Q_brief}) and (\ref{J_meso}), which is traceless by definition, we rewrite  Eq.~(\ref{J_mic_conservation}) using ${\sf J}^{mic}$ of Eq.~(\ref{J_traceless_mic}),
\begin{equation}
	\partial_i\partial_j {J}^{mic}_{ij} + \textstyle{1\over 2}\nabla^2\rho^{mic} =
		\nabla\cdot \delta {\bf t},
\end{equation}
and finally write down the mesoscopic tensor conservation law:
\begin{equation}
	\partial_i\partial_j J_{ij}+\textstyle{1\over 2}\ell_0\nabla^2\rho = 
		\nabla\cdot{\bf g},
		\label{tensorial_continuity}
\end{equation}
where $J_{ij}({\bf x}) = \rho({\bf x})\ell_0 Q_{ij}({\bf x})$.  The tensorial form of the conservation law in Eq.~(\ref{tensorial_continuity}) for quadrupolar ordering should be contrasted with Eq.~(\ref{current_continuity}) in the case of polar ordering. 
In the absence of polar ordering or in the infinite chain limit ${\bf g}=0$ and Eq.~(\ref{tensorial_continuity})
reduces to 
\begin{equation}
	\partial_i\partial_j J_{ij}+\textstyle{1\over 2}\ell_0\nabla^2\rho = 0.
	\label{tensorial_continuity_nosources}	
\end{equation}
The tensorial conservation law, Eq.~(\ref{tensorial_continuity}), presents a constraint that any configuration of a flexible nematic polymer must satisfy. Just as the polar constraint in Eq.~(\ref{current_continuity}) introduces a coupling between the orientational order parameter and density variation, the tensorial constraint connects the  $\sf Q$-tensor gradients with the density variations. Not unlike simulations of polar polymers~\cite{svensek-podgornik,svensek-podgornik_chiral,klug}, the constraint of Eq.~(\ref{tensorial_continuity_nosources}) would need to be enforced in coarse-grained models of nematic polymers that are based on the nematic $\sf Q$-tensor.

For small deviations from a homogeneous configuration one can derive a lowest order coupling between the deformations in the density and orientational fields.  Let us linearize the constraint of Eq.~(\ref{tensorial_continuity_nosources}) for the case where deviations from homogeneous director and density fields keep the nematic ordering uniaxial and its degree $S$ fixed. Assuming 
\begin{equation}
	{\bf n}=  \hat{\bf e}_z + \delta {\bf n}({\bf r}), \quad \delta{\bf n}=(\delta n_x, \delta n_y), \quad \rho = \rho_0 + \delta\rho({\bf r})
\end{equation}
and the form  $	Q_{ij} = {3\over 2}S (n_i n_j - {1\over 3}\delta_{ij})$ for the nematic $\sf Q$-tensor, the linearization of the tensorial constrain then leads to 
\begin{equation}
	(S+{\textstyle{1\over 2}})\,\partial_z^2 \delta\rho + {\textstyle{1\over 2}}(1-S)\, \nabla\!_\perp^2 \delta\rho + {\textstyle{3}} S\rho_0\, \partial_z \left(\nabla\!_\perp \cdot \delta{\bf n}\right) = 0,
	\label{linearized_tenzor}
\end{equation}
where $\nabla\!_\perp = (\partial/\partial x, \partial/\partial y)$. This is to be contrasted with the analogous linearized constraint in the vectorial case,
\begin{equation}
	{\partial\delta\rho\over\partial z} + \rho_0 a\, \nabla\!_\perp\cdot\delta{\bf n} = 0,
	\label{linearized_vector}
\end{equation}
with ${\bf a}= a{\bf n}$. 
The latter constraint implies that the splay deformation of the director induces a density gradient parallel to the director.

Unlike small molecule nematics where density is decoupled from orientational order, the connectivity of the nematic polymer is reflected in the vectorial (Eqs.~(\ref{current_continuity_nosources}) and (\ref{linearized_vector})) or in the tensorial constraint, Eqs.~(\ref{tensorial_continuity}) and (\ref{linearized_tenzor}). 
In the quadrupolar nematic phase, where hairpins may be present or not, Fig.~\ref{connectivity_polarity}, the vectorial and tensorial constraints represent two limits. In the rigid chain limit, i.e., if the persistence length of the polymer is large compared to the mesoscopic length ($V^{1/3}$, where $V$ is the coarse-graining volume) and the mesoscopic volume is free of hairpins it is possible to unambiguously, but arbitrarily, assign polar chain orientations so that the ordering of chain tangents may be described as polar, and the corresponding vectorial constraint is valid. In the flexible chain limit where the persistence length is small, the hairpins are abundant and it is not possible to uniformly orient tangents, their ordering is necessarily quadrupolar and hence the tensorial constraint holds.

\section{Hairpins and the tensorial order}

\label{hairpins}

\noindent
The existence of hairpin turns \cite{kamien} in polymer nematics underlies a key microscopic distinction between (polar) vectorial and (quadrupolar) tensorial order. In the derivation of Sec.~\ref{derivation}, there is no explicit constraint on hairpin number or distribution. Intuitively, on expects that in a physical system, the abundance of hairpins is determined by the ratio of the mesoscopic (coarse-graining) volume length and the persistence length of the chain.  In this section, we argue that in general hairpin distributions influence {\it gradients} of $Q_{ij}$, and under certain restricted conditions, the hairpin distribution is encoded entirely by mesoscopic order parameter variations.

\begin{figure}[htb]
\begin{center}
\includegraphics[height=40mm]{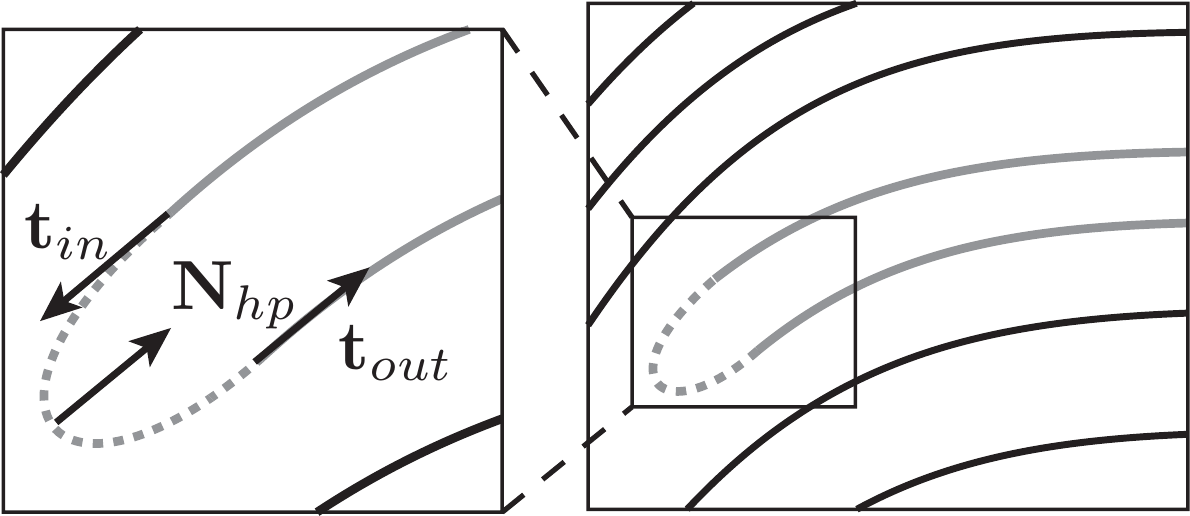}
	\caption{Schematics of hairpin geometry with a (locally) straight chain segment and a hairpin, idealized.  ${\bf N}^{hp}$ labels the orientation of the hairpin (dashed segment), and points normal to the chain at maximal curvature. The hairpin orientation is invariant under flip of the chain direction (${\bf t}(s) \longrightarrow -{\bf t}(s)$). In the case of a point-like hairpin, the tensor order parameter does not distinguish between the straight chain and the hairpin. The vector order parameter of course does, being finite in the first case and zero in the second. \label{hairpin}	}
\end{center}
\end{figure}

It is straightforward to understand the destructive influence of hairpins on polar order by rewriting the definition of the vector order parameter $\bf a$, Eq.~(\ref{j_meso}), considering for simplicity only a section of the chain $s_1<s<s_2$ inside $V({\bf x})$,
\begin{equation}
	{\bf a} = {1\over L}\int_{{\bf x}(s)\in V({\bf x})} ds ~{d{\bf x}(s)\over ds}({\bf x}(s))=
			{1\over L} \left[{\bf x}(s_2)-{\bf x}(s_1)\right].
\end{equation}
It is obvious that $\bf a$ depends only on the end and starting points and the length $L$ of the considered section, but not on its conformation in between. In particular, bringing the end two points together, e.g., in case of a (nonideal) hairpin, the polar ordering is identically zero, irrespective of the actual curvature profile of the chain section.

The tensor order parameter, however, is little affected by the hairpins, and in the limit of point-like turns not at all. Rewriting Eq.~(\ref{J_meso}), again for a chain section of length $L$ inside the mesoscopic volume $V({\bf x})$,
\begin{eqnarray}
	Q_{ij} &=& {1\over L}\int_{{\bf x}(s)\in V({\bf x})} ds ~{\textstyle{3\over 2}}\left[{d x_i(s)\over ds} {d x_j(s)\over ds}-{\textstyle{1\over 3}}\delta_{ij}\right] 
	=\nonumber\\
	&=& {3\over 2}{1\over L}\left[{dx_i\over ds}(s_2)\, x_j(s_2)-{dx_i\over ds}(s_1)\, x_j(s_1)
		-\int_{{\bf x}(s)\in V({\bf x})} ds ~{d^2x_i(s)\over ds^2}\,x_j(s)\right]-		
		{1\over 2}\delta_{ij},
		\label{Q_curvature}
\end{eqnarray}
one discerns that besides being dependent on the positions of the end and the starting points of the chain section, $\sf Q$  depends also on the chain tangents in these points and on the curvature profile of the whole section.  Due to the vanishing arclength of hairpins, it can be easily verified that for a straight hairpin with a point-like U-turn, Eq.~(\ref{Q_curvature}) gives a uniaxial $\sf Q$, Eq.~(\ref{Q_uniaxial}), with the degree of ordering $S=1$.  We note again, therefore, that the $\sf Q$ tensor, describing the degree of uniaxial ordering and the biaxiality, cannot capture the density of hairpins.  In the limit of point-like turns, $\sf Q$ is not affected by the hairpins at all.  Notwithstanding the insensitivity of local value of $\sf Q$ to hairpins, such configurations are associated with an energy cost.   In order to take this energy into account in a {\it continuum} description, an extension to the nematic order parameter would be needed for polymer nematics, describing the magnitude of the microscopic curvature of the chain.


An illuminating way to single out the effect of hairpins is to consider the extreme case of uniform density and perfect order. In this case, as it will turn out, the contribution of singular hairpins can be isolated by considering the gradients of the $\sf Q$ tensor. In order to do that let us introduce the vector field, defined microscopically as
\begin{equation}
{\kappa_i}^{mic}({\bf x}) = \frac23 \partial_j \tilde{J}^{mic}_{ij}({\bf x}) = \int_{{\bf x}(s)}ds~  t_i(s) t_j(s)~  \partial_j \delta({\bf x} - {\bf x}(s)).
\end{equation}
After the integration by parts and application of the Frenet-Serret formulas~\cite{kamien_rmp} this can be reduced to
\begin{equation}
\pmb{\kappa}^{mic}({\bf x}) = - \delta {\bf t}({\bf x}) + \int_{{\bf x}(s)}ds~  \kappa(s) {\bf N}(s) ~ \delta({\bf x} - {\bf x}(s)),
\label{kappamic}
\end{equation}
where $\kappa(s)$ and ${\bf N}(s)$ are the respective curvature and unit normal of the chain ${\bf x}(s)$.  As above, in the absence of polar order, the average of $\delta {\bf t}({\bf x})$ necessarily vanishes, which implies that coarse-graining $\pmb{\kappa}^{mic}$ gives local mesoscopic average of chain curvature, $\kappa(s) {\bf N}(s)$.  We note, in passing, that the introduction of the curvature field, allows one reformulate the tensorial conservation law derived above in terms of conserved vector quantity, $\nabla \cdot \pmb{\kappa} = 0$.  

A point-like hairpin can be defined as a singular region of chain bending which does not alter the mean orientation or segment density within a coarse-grained volume, Fig.~\ref{hairpin}. In the limit of vanishing hairpin length, we must have $\kappa(s) {\bf N} (s) \longrightarrow 2 {\bf N}^{hp} \delta( s - s^{hp}) + \kappa^{ns}(s) {\bf N}^{ns}(s)$, where ${\bf N}^{hp}$ is the orientation of a hairpin normal located at $s^{hp}$ and $\kappa^{ns}(s)$ and ${\bf N}^{ns}(s)$ are the curvature and normal of the non-singular (smooth) lengths of the chain that dominate the local orientational order. For a chain possessing a general distribution of point-like hairpins of orientation ${\bf N}^{\alpha}$ at locations $s_{\alpha}$, microscopic chain curvatures, Eq.~(\ref{kappamic}), can be decomposed in the same manner which upon coarse-graining yields the mesoscopic bending field as
\begin{equation}
\pmb{\kappa} ({\bf x}) = {\bf h}({\bf x})+ \pmb{\kappa}^{ns}({\bf x}),
\end{equation}
where $\kappa^{ns}({\bf x})$ is the average chain-bending field {\em excluding} hairpins and ${\bf h}({\bf x})$ is the {\it hairpin field} defined by 
\begin{equation}
{\bf h}({\bf x}) =\frac{2}{V} \sum_{{\bf x}_{\alpha} \in V({\bf x})} {\bf N}^{\alpha},
\end{equation}
${\bf N}^{\alpha}$ and ${\bf x}_{\alpha}$ are the respective orientation and position of hairpins.

\begin{figure}[t]
\begin{center}
	\includegraphics[width=90mm]{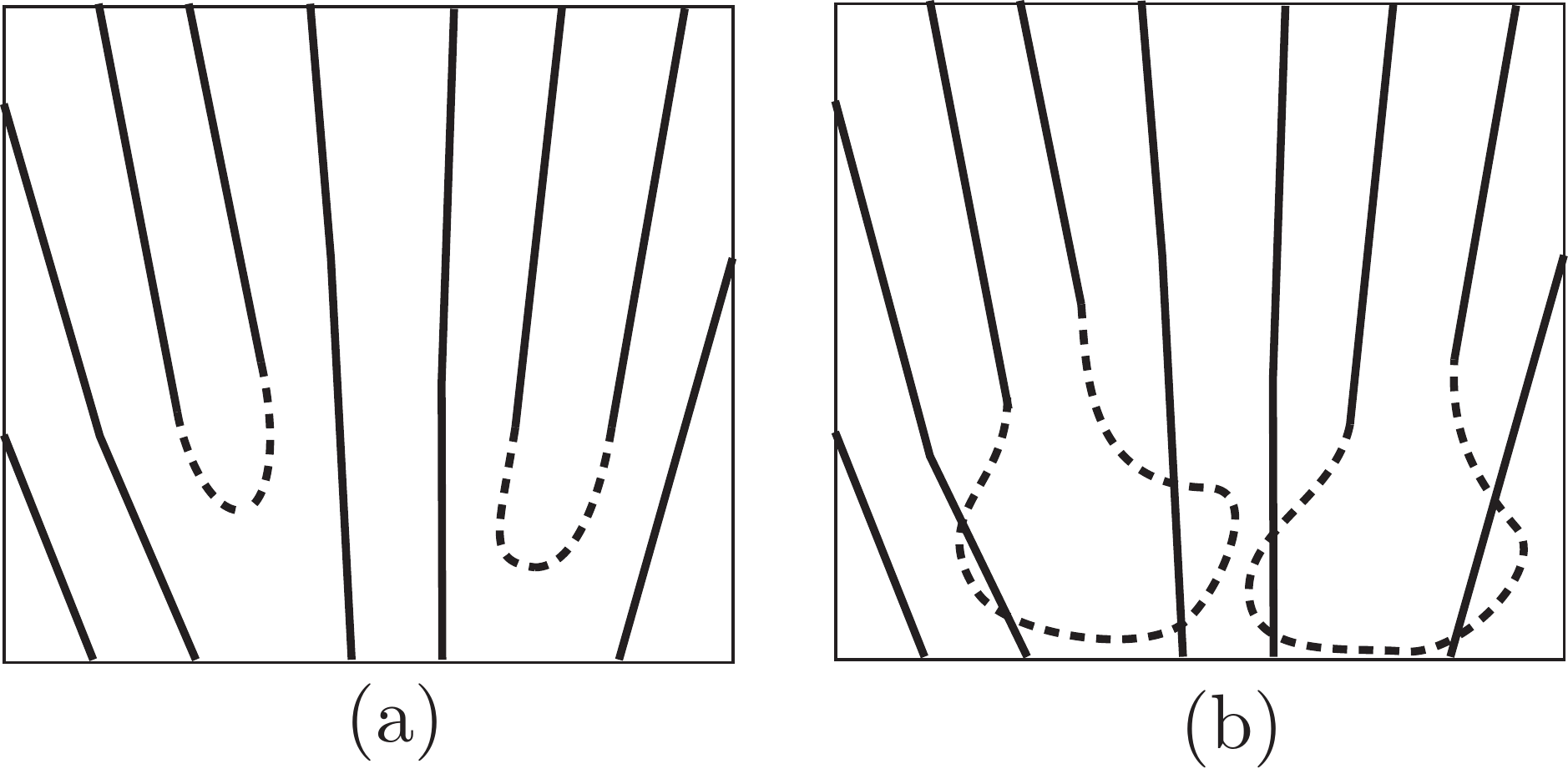}
	\caption{In (a), schematic of hairpins (dotted) as infinitesimal length, $\pm 2$ sources or sinks of chain segments, accommodating director splay at constant chain density. In (b), finite-length loops accommodating an equivalent director splay as in (a).
	\label{newchain_hairpins}
	}
\end{center}
\end{figure}

From the perspective of the local density, order parameter and director fields, $\rho({\bf x})$, $S({\bf x})$ and ${\bf n}({\bf x})$, respectively, it is straightforward to show that the mesoscopic bending field is given by
\begin{equation}
\pmb{\kappa}= {\bf n} ({\bf n} \cdot \nabla)(\rho \ell_0 S) + \frac13 \nabla  \Big[( \rho \ell_0 (1 - S)\Big] + \rho \ell_0 S\Big[ {\bf n} (\nabla \cdot {\bf n})+ ({\bf n} \cdot \nabla) {\bf n}\Big].
\end{equation}
Focusing on the extreme case of uniform density and perfect order ($\rho = {\rm const.}$, $S = 1$), we notice that $\pmb{\kappa}({\bf x})$ includes two components that are parallel and perpendicular to the local director, respectively. The latter contribution, $\rho \ell_0 S  ({\bf n} \cdot \nabla) {\bf n}$, represents the curvature of the director field itself, and hence, its square is a familiar representation of the bending energy density in the Frank elastic energy. For the case of perfect, uniform density order, all lengths contributing to $\sf Q$ tensor are in perfect alignment with ${\bf n}$, hence the normal to non-singular spans, ${\bf N}^{ns}(s)$, is uniformly perpendicular to the director. Likewise, as all hairpins are joined to uniformly oriented non-singular segments, the hairpin normals lie {\it parallel} to the non-singular segments, contributing only in the ${\bf n}$ direction. Therefore, in the case of uniform density and perfect orientational order we can unambiguously make the identification
\begin{equation}
\rho \ell_0 S {\bf n}  (\nabla \cdot {\bf n} )  = {\bf h} ({\bf x})  \qquad {\rm and} \qquad  \rho \ell_0 S({\bf n} \cdot \nabla) {\bf n}= \pmb{\kappa}^{ns}({\bf x}). 
\end{equation}
Thus, we can directly relate the splay of the director field to the local density of hairpins~\cite{selinger_bruinsma_jphys}, which act as $\pm 2$ sources and sinks of smoothly configured chains (see Fig.~\ref{newchain_hairpins}a).

In the more general case of arbitrary variations in density and order parameter it is {\sl not possible} to decouple the contributions of hairpins and finite-length loops, both of which facilitate splay (see Fig.~\ref{newchain_hairpins}b). Nevertheless, as a measure of the average chain curvature, $\pmb{\kappa}({\bf x})$ retains considerably more microscopic information than the curvature of the mean orientation field, $\rho \ell_0 S ({\bf n} \cdot \nabla) {\bf n}$. 

\section{Vectorial vs. tensorial constraint}

\label{vecvstens}

\noindent
Here, we compare the consequences of the tensorial conservation law, Eq.~(\ref{tensorial_continuity_nosources}), and the vectorial one, Eq.~(\ref{current_continuity_nosources}), in the simplest limit. We assume that the degree of the nematic order and the density are fixed, corresponding to the limit of weak director field deformations that are in addition such as to keep the density constant. Using the uniaxial $\sf Q$-tensor {\it Ansatz} Eq.~(\ref{Q_uniaxial}) in Eq.~(\ref{tensorial_continuity_nosources}) and setting 
$\nabla S=\nabla \rho=0$, one gets the minimal form of the tensorial conservation law:
\begin{equation}
	\partial_i\partial_j n_i n_j = 
		{\nabla}\cdot\Big[{\bf n}(\nabla\cdot{\bf n})+({\bf n}\cdot\nabla){\bf n}\Big]=0,
		\label{tensorial_continuity_simple}
\end{equation}
as a contrast to the vectorial conservation law in the same limit, which requires simply the absence of splay,
\begin{equation}
	{\nabla}\cdot{\bf n}  = 0.
	\label{div_zero}
\end{equation}
Thus, already in this limit the constraints are quite different.

We can find a simple solution of Eq.~(\ref{tensorial_continuity_simple}) with cylindrical symmetry, in polar coordinates $(r,\phi)$: 
\begin{equation}
	{\bf n} = \hat{\bf e}_r \cos\psi(r) + \hat{\bf e}_\phi \sin\psi(r),
	\label{ansatz}
\end{equation}
where $\psi$ is the angle between $\hat{\bf e}_r$ and $\bf n$.  Because the $\phi$-derivative of any vector component is zero due to the cylindrical symmetry of the ansatz, 
the divergence in Eq.~(\ref{tensorial_continuity_simple}) 
involves only the radial component of the vector ${\bf n}(\nabla\cdot{\bf n})+({\bf n}\cdot\nabla){\bf n}$.
Then we have
\begin{eqnarray}
	\hat{\bf e}_r \cdot	{\bf n}({\nabla\cdot{\bf n}}) &=& \cos\psi {1\over r}{\partial\over\partial r}(r\cos\psi),\\
	{\bf\hat{e}}_r \cdot ({\bf n}\cdot\nabla){\bf n} &=& \left(\cos\psi {\partial\cos\psi\over\partial r}-{\sin^2\psi\over r}\right)
\end{eqnarray}
and so Eq.~(\ref{tensorial_continuity_simple}) assumes the form
\begin{equation}
	{1\over r}{\partial\over\partial r}\left[\cos\psi{\partial\over\partial r}(r\cos\psi)
		+ r\cos\psi{\partial\cos\psi\over\partial r}-\sin^2\psi\right]  = 0.
\end{equation}
Integrating and writing the second term in the form of the first one, one finds the solutions
\begin{equation}
	\cos^2\psi = A + {B\over r^2},
\end{equation}
where $A$ and $B$ are constants. 
The divergence of this director field is 
\begin{equation}
	\nabla\cdot{\bf n} = {1\over r}{\partial\over\partial r}(r\cos\psi)=\pm{A\over\sqrt{A r^2+B}},
\end{equation}
which is nonzero if $A\ne 0$. This contrasts with the vectorial constraint, Eq.~(\ref{div_zero}).

\begin{figure}[htb]
\begin{center}
	\mbox{
		\subfigure[\, $A=0$, $B=1$]{\includegraphics[width=35mm]{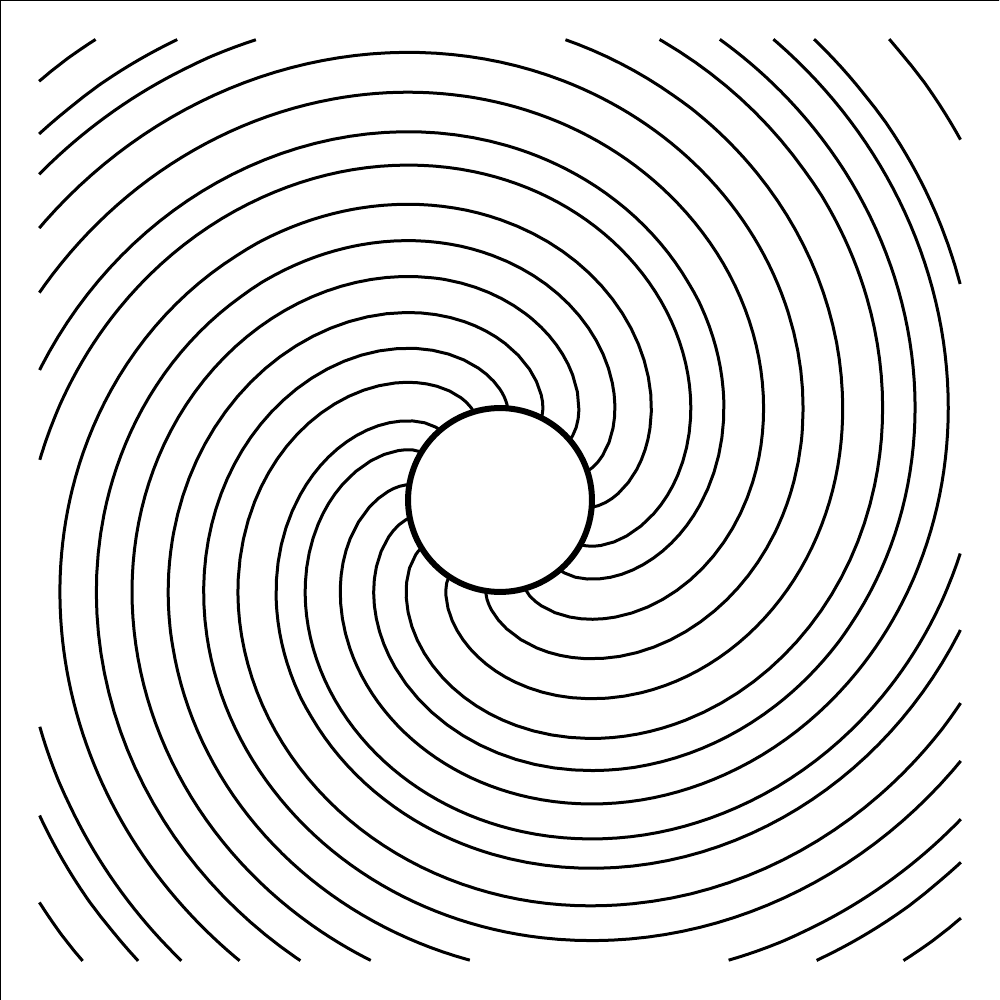}}\hspace{5mm}  
		\subfigure[\, $A=1$, $B=-1$]{\includegraphics[width=35mm]{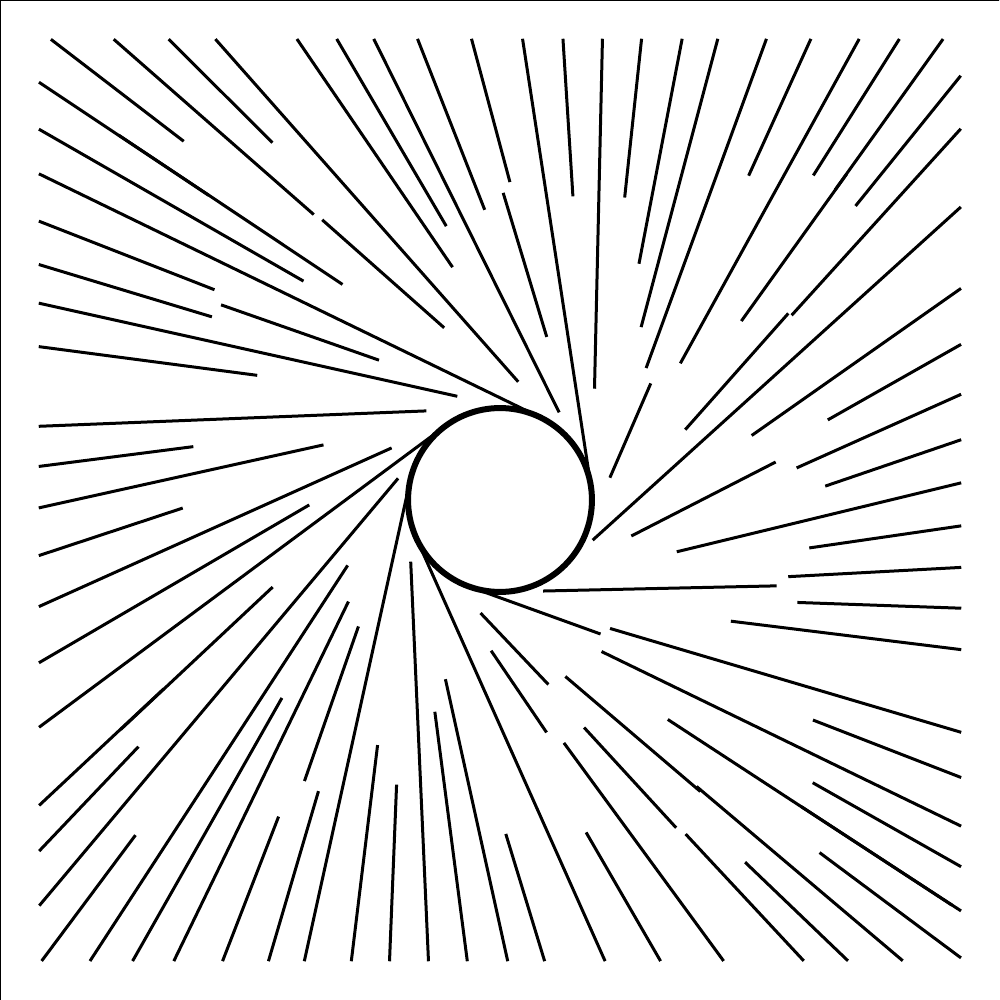}}\hspace{5mm}  
		\subfigure[\, $A=0.5$, $B=0.5$]{\includegraphics[width=35mm]{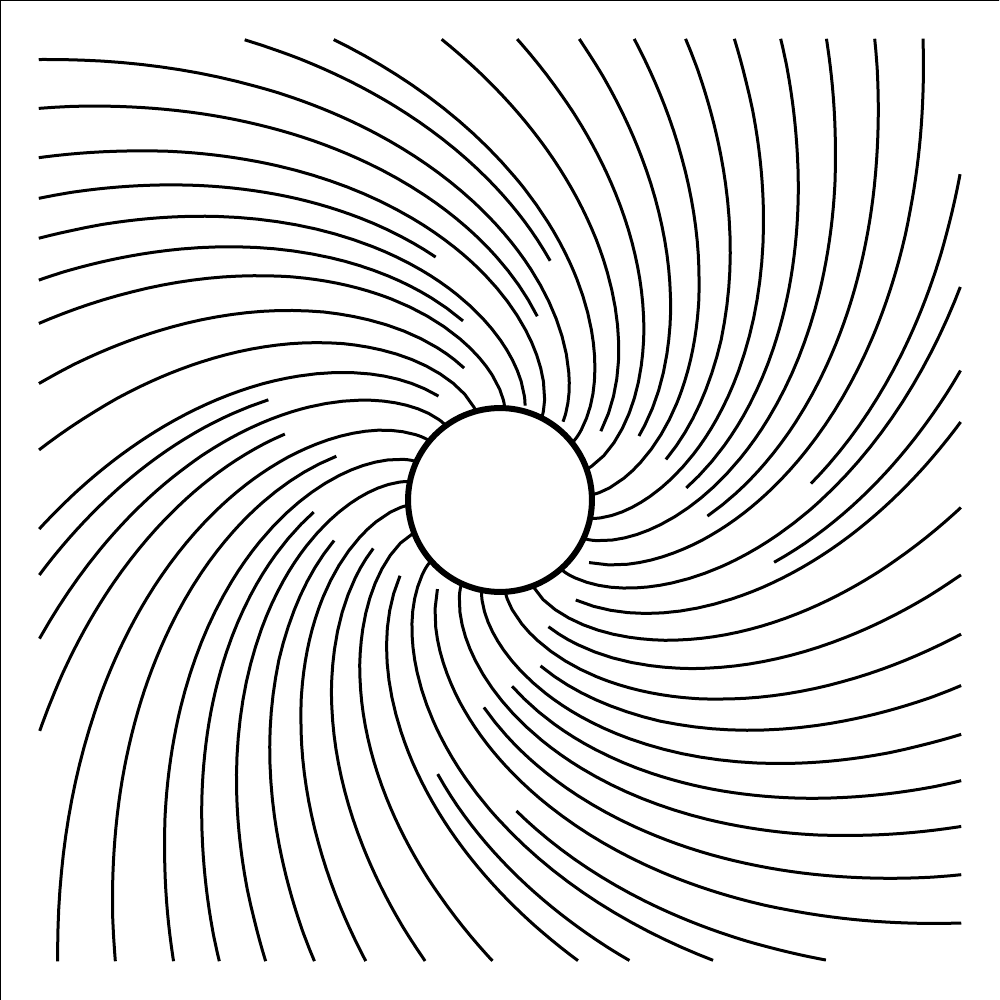}}\hspace{5mm}  
		\subfigure[\, $A=0.5$, $B=-0.5$]{\includegraphics[width=35mm]{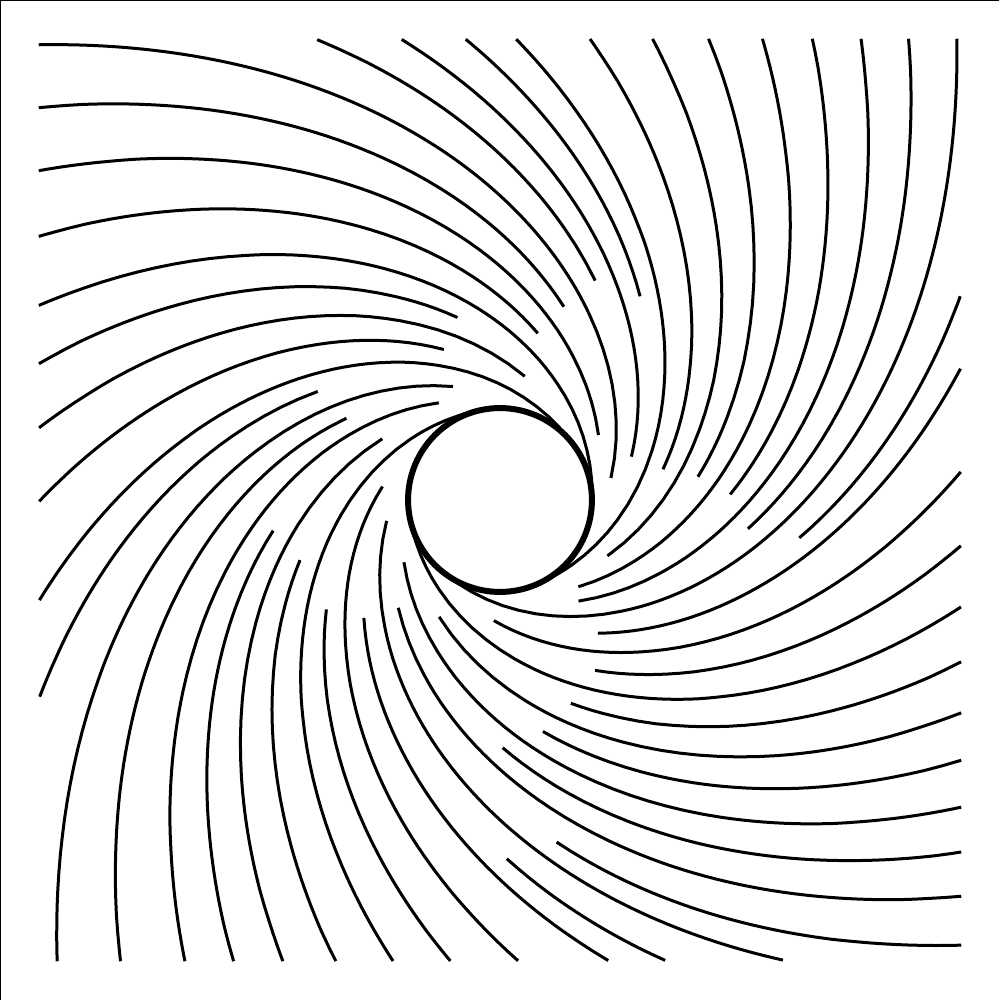}}
	}
	\caption{Director field streamlines of some solutions (\ref{spiral_infinity}) of the constraint (\ref{tensorial_continuity_simple}): (a) if $\bf n$ is tangential at $r\to\infty$, the configuration is divergence-free for any $\psi_1$, whereas (b) if $\bf n$ is radial at $r\to\infty$, there is no bend deformation for any $\psi_1$ --- both is revealed nicely by the streamlines. In (c) and (d), $\psi_\infty=45^\circ$; $\psi$ can either increase (c) or decrease (d) with $r$.
	\label{spirals}
	}
\end{center}
\end{figure}

Generally, let the configuration be defined for $r_1<r<r_2$ and $\psi(r_1)=\psi_1$, $\psi(r_2)=\psi_2$, then 
\begin{eqnarray}
	A&=&{(r_2\cos\psi_2)^2-(r_1\cos\psi_1)^2\over r_2^2-r_1^2},\\
	B&=&{(r_1 r_2)^2\over r_2^2-r_1^2}(\cos^2\psi_1-\cos^2\psi_2).
\end{eqnarray}
In the special case when $r_2\to\infty$, putting $\psi_2=\psi_\infty$, we have $ A =  \cos^2\psi_\infty$ and $ B =  r_1^2 (\cos^2\psi_1-\cos^2\psi_\infty)$ so that the solution is
\begin{equation}
	\cos^2\psi = \cos^2\psi_\infty + (\cos^2\psi_1-\cos^2\psi_\infty) {r_1^2\over r^2}
	\label{spiral_infinity}
\end{equation}
and the divergence of the director field is nonzero if $\cos\psi_\infty\ne 0$, i.e., if the director at $r\to\infty$ is not tangential. Some of the solutions in Eq.~(\ref{spiral_infinity}) are presented in Fig.~\ref{spirals}.

\begin{figure}[htb]
\begin{center}
	\mbox{
		\subfigure[\, $A=0$, $B=0$]{\includegraphics[width=35mm]{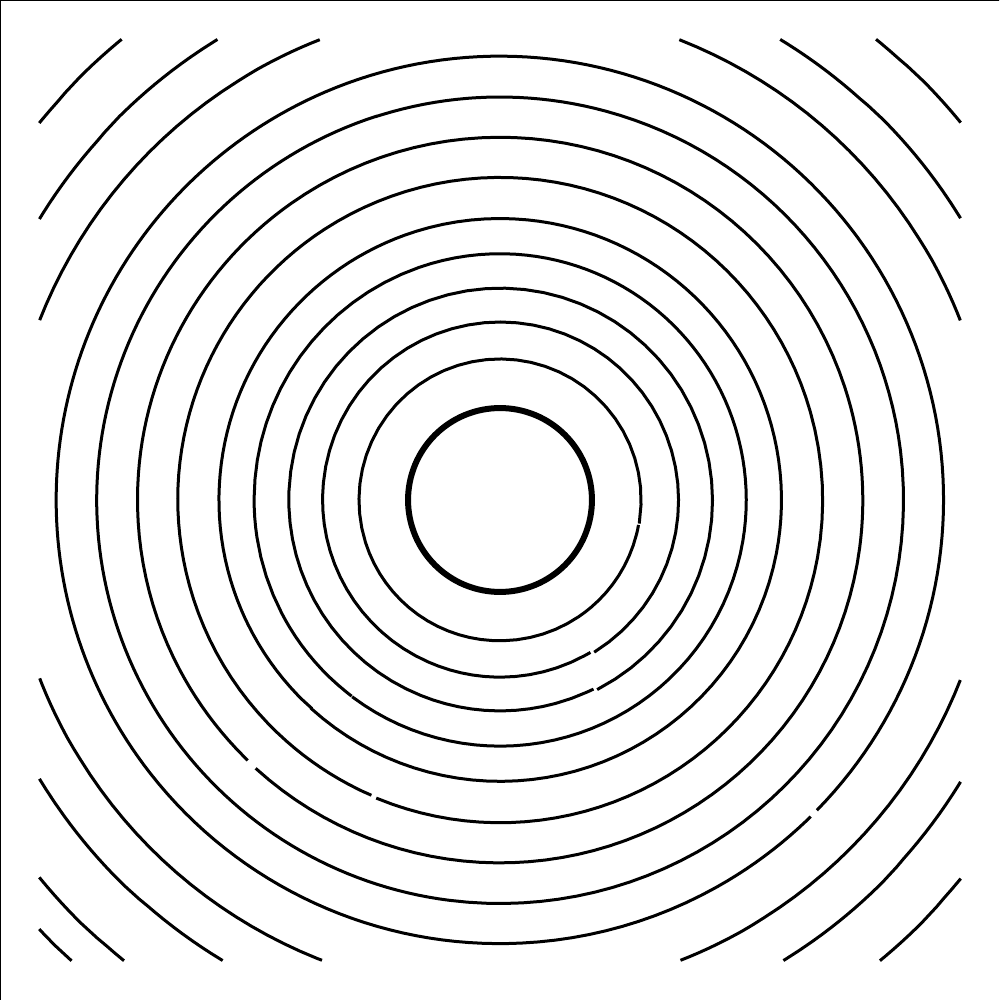}}\hspace{5mm}  
		\subfigure[\, $A=1$, $B=0$]{\includegraphics[width=35mm]{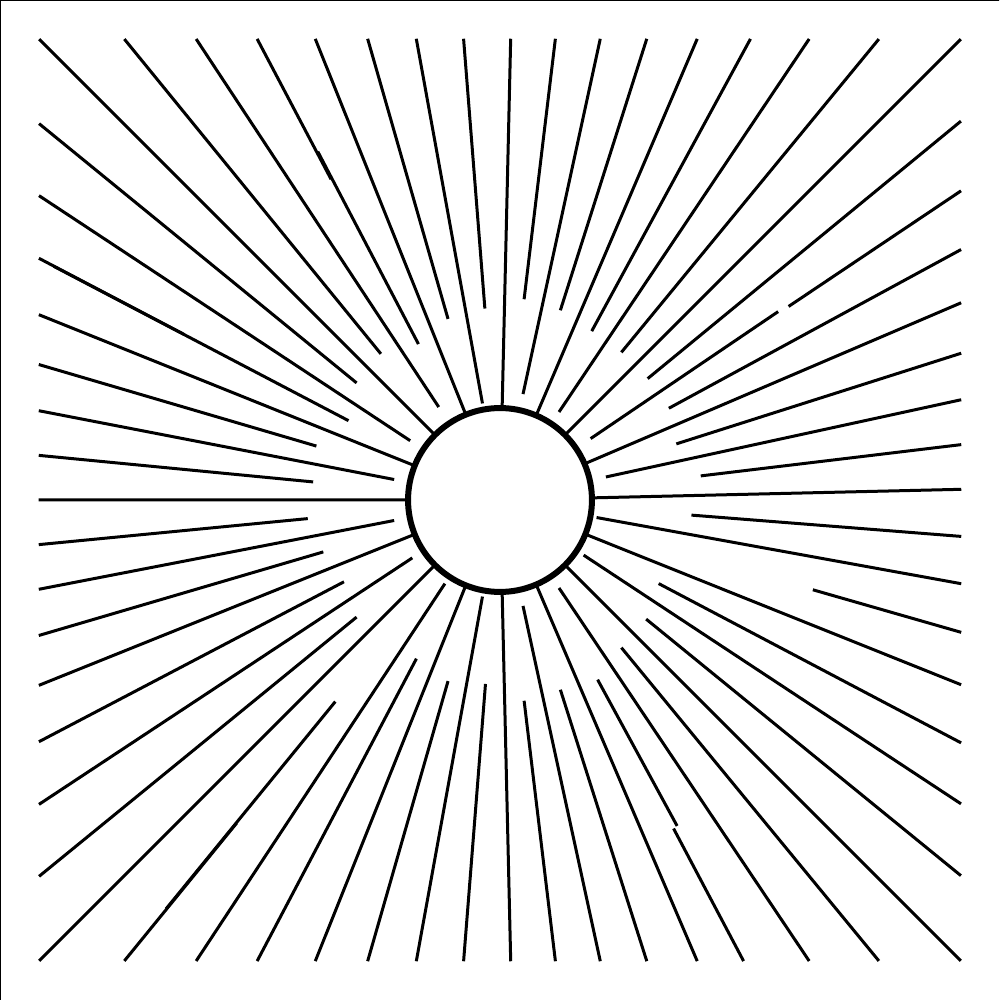}}\hspace{5mm}  
		\subfigure[\, $A=0.026$, $B=0$]{\includegraphics[width=35mm]{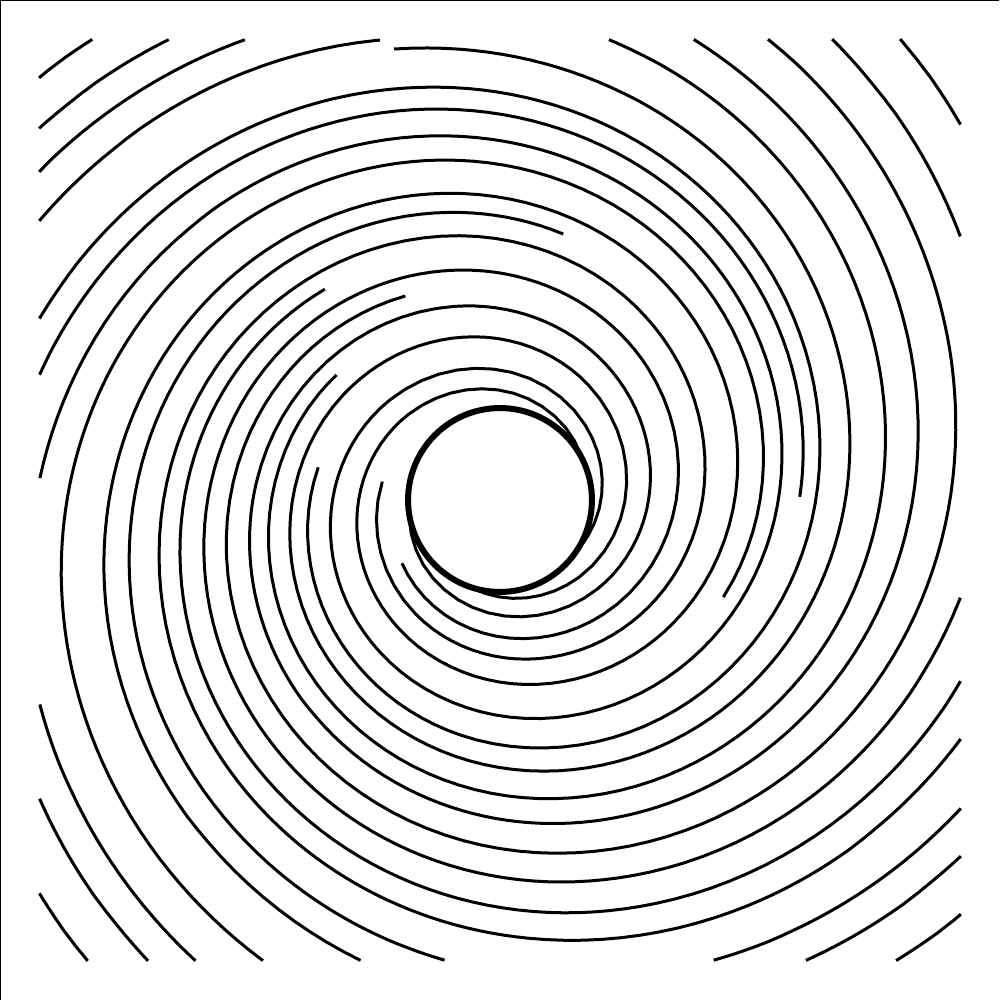}}\hspace{5mm}  
		\subfigure[\, $A=0.5$, $B=0$]{\includegraphics[width=35mm]{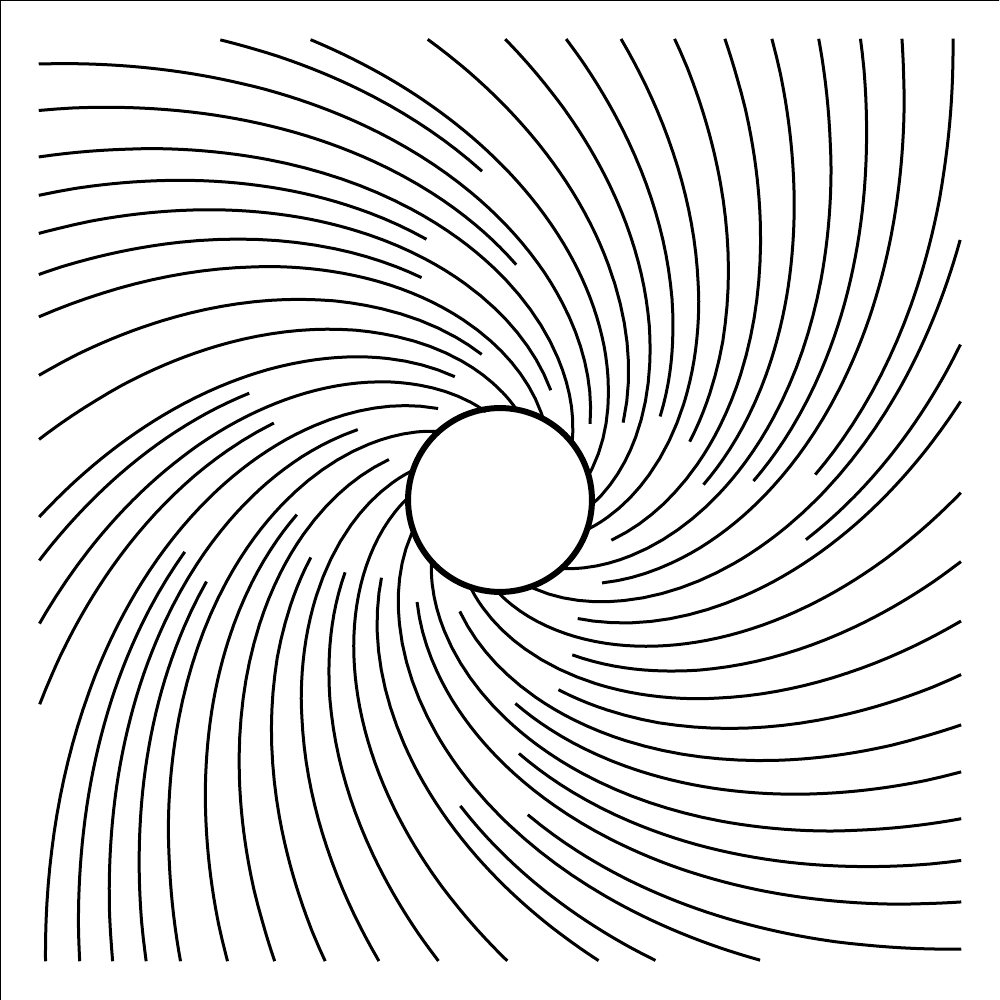}}
	}
	\caption{Director field streamlines of some configurations with $\psi={\rm const}$, which satisfy the constraint (\ref{tensorial_continuity_simple}) and at the same time minimize the one-elastic-constant free energy. Only the circular configuration (a) has no splay. The coil-like configuration (c) is splayed only weakly.
	\label{Fminimizers}
	}
\end{center}
\end{figure}

Let us relate this solution to the free energy minimum in one elastic constant approximation, which satisfies
\begin{equation}
	{\bf n}\cdot \nabla^2 {\bf n} = 0, \qquad {\rm or} \qquad \nabla^2 \theta =0,
\end{equation}
with the parametrization ${\bf n}=\hat{{\bf e}}_x \cos\theta + \hat{{\bf e}}_y \sin\theta$. The solution of the above equation with cylindrical symmetry as before, thus with $\psi(r) = \theta(r) - \phi$, is
\begin{equation}
	\psi(r)=C\ln{r\over r_1} + D,	
\end{equation}
where $C$ and $D$ are constants.  Hence, the family of solutions with the cylindrical symmetry that both satisfy the tensorial conservation law, Eq.~(\ref{tensorial_continuity_simple}), and minimize the one elastic constant free energy at the same time is
\begin{equation}
	\psi = {\rm const.},
\end{equation}
Fig.~\ref{Fminimizers}.
Among these, only the solution $\psi=\pi/2$ (the circular disclination) has no splay and is thus the only allowed in the vectorial case. In contrast, the tensorial conservation law allows solutions with an arbitrary constant $\psi$. Coil-like solutions with $\psi$ close to $\pi/2$, Fig.~\ref{Fminimizers}c, are easily imagined to be formed by a single polymer chain. One has to bear in mind though that the figures show the director field, not the chain(s) --- these must make hairpins microscopically to accommodate the splay at the constant density.  In light of the discussion of the previous section, in the extreme case of uniform, perfect order, the local concentration and orientation of hairpins  is directly proportional to local splay, ${\bf n} (\nabla \cdot {\bf n})$.

\section{Conclusions and discussion}
\label{conclusion}

\noindent
We have derived the appropriate constraint for the (mesoscopic) nematic tensor field valid in the case of nematic polymers. The constraint bares some resemblance to the case of a polar order but is in general fundamentally different. The two descriptions differ most clearly in their treatment of the hairpin configurations. While the tensor order parameter does not distinguish between the straight chain and the hairpin, the vector order parameter does, being finite in the first case and zero in the second.  

The new tensorial order parameter constraint should be included in all mesoscopic descriptions of nematic ordering of stiff polymer chains such as DNA that are known to make nematic phases and cannot be captured in the polar description. In particular, boundary conditions necessarily introduce topological defects into the configuration of spherically-confined polymer nematics~\cite{shin}.  Since a vector order parameter cannot describe the full defect spectrum of a nematic (disclinations of charge $\pm 1/2$, $\pm 1$, ...), it is expected that equilibria of vector and tensor models of confined polymers exhibit markedly distinct topologies. The full consequences of this new constraint in particular in the case of confined nematic polymer liquid crystals remain to be explored.  

The derived constraint for the (mesoscopic) nematic tensor field also opens up a full Landau-de Gennes description of polymer ordering based on an analogy with simple nematogens. The free energy {\sl Ansatz}~ is in fact the same but among all the possible solutions only those that satisfy the new constraints are physically admissible. Furthermore we found out that as a measure of the average chain curvature, the mesoscopic bending field retains considerably more microscopic information than the curvature of the mean orientational field. Order parameters and their invariants  based on the mesoscopic bending field rather than the standard Frank elastic form could capture the energetics of microscopic chain configurations, like singular hairpins, underlying the mesoscopic order of nematic polymers. We intend to pursue these lines of thought in a subsequent publication.

\begin{acknowledgments}
DS and RP acknowledge the support of the Agency for Research and Development of Slovenia (Grants No. J1-4297, J1-4134).
GG acknowledges funding support from NSF CAREER Award DMR 09-55760 and the Alfred P. Sloan Foundation.
\end{acknowledgments}


\begin{thebibliography}{99}

\bibitem{meyer} V. G. Taratuta and R. B. Meyer, {\sl Liquid Cryst.} {\bf 2} (1987) 373.
\bibitem{kamien}R. D. Kamien, P. Le Doussal, and D. R. Nelson, {Phys. Rev. A} {\bf 45} 8727 (1992).
\bibitem{Kamien2} R. D. Kamien and T.C. Lubensky, {\sl J. Phys. I France} {\bf 3} (1993) 2131.
\bibitem{conteq} P. G. de Gennes, {\sl Mol. Cryst. Liq. Cryst. Lett. } {\bf 34} 177 (1977).
\bibitem{meyer1} R. Meyer in {\sl Polymer Liquid Crystals}, A. Ciferri et. al. (eds.), Academic NY (1982).
\bibitem{selinger_bruinsma_pra} J. V. Selinger and R. F. Bruinsma, {\sl Phys. Rev. A} {\bf 43} 2910 (1991).
\bibitem{kamien_toner} R. D. Kamien and J. Toner, {\sl Phys. Rev. Lett.} {\bf 74} 3181 (1995).
\bibitem{svensek-podgornik} D. Sven\v sek, G. Veble, and R. Podgornik, {\sl Phys. Rev. E} {\bf 82} (2010) 011708.
\bibitem{svensek-podgornik_chiral} D. Sven\v sek and R. Podgornik, {\sl Europhys. Lett.} {\bf 100} (2012) 66005.
\bibitem{degennes_prost} P.-G. de Gennes and J. Prost {\sl The Physics of Liquid Crystals}, 2nd ed., Oxford NY (1993).
\bibitem{klug} W.S. Klug and M. Ortiz, {\sl J. Mech. Phys. Solids} {\bf 51} 1815 (2003).
\bibitem{kamien_rmp} R. D. Kamien, {\sl Rev. Mod. Phys.} {\bf 74} 953 (2002).
\bibitem{selinger_bruinsma_jphys} J. V. Selinger and R. F. Bruinsma, {\sl J. Phys. II France} {\bf 2} 1215 (1992).
\bibitem{shin} H. Shin and G. M. Grason, {\sl Europhys. Lett.} {\bf 96} 36007 (2011).

\end{thebibliography}
\end{document}